# A New Parallelization Method for K-means

Shikai Jin, Yuxuan Cui, Chunli Yu

**Abstract** - K-means is a popular clustering method used in data mining area. To work with large datasets, researchers propose PKMeans, which is a parallel k-means on MapReduce [3]. However, the existing k-means parallelization methods including PKMeans have many limitations. It can't finish all its iterations in one MapReduce job, so it has to repeat cascading MapReduce jobs in a loop until convergence. On the most popular MapReduce platform, Hadoop, every MapReduce job introduces significant I/O overheads and extra execution time at stages of job start-up and shuffling [2]. Even worse, it has been proved that in the worst case, k-means needs $2^{\Omega(n)}$ MapReduce jobs to converge [4, 5], where n is the number of data instances, which means huge overheads for large datasets. Additionally, in PKMeans, at most one reducer can be assigned to and update each centroid, so PKMeans can only make use of limited number of parallel reducers.

In this paper, we propose an improved parallel method for k-means, IPKMeans, which has a parallel preprocessing stage using k-d tree [8] and can finish k-means in one single MapReduce job with much more reducers working in parallel and lower I/O overheads than PKMeans and has a fast post-processing stage generating the final result. In our method, both k-d tree and the new improved parallel k-means are implemented using MapReduce and tested on Hadoop. Our experiments show that with same dataset and initial centroids, our method has up to 2/3 lower I/O overheads and consumes less amount of time than PKMeans to get a very close clustering result.

## 1. Introduction

### (i) MapReduce

Big data is a hot topic in cloud computing area. To deal with the increasing amount of data, MapReduce is designed to process and generate large datasets with parallel, distributed algorithm on a cluster. Many single-machine algorithms have been successfully adapted and implemented using MapReduce to gain significant speed-up. However, there are overheads during the of one MapReduce job. Let's go through the execution steps of one MapReduce job to see where overheads come from [1].

1. The MapReduce library in the user program first splits the input files into M pieces.
2. The master node picks idle workers and assign each one a map task or a reduce task. Let's assume we have M map tasks and R reduce tasks.
3. A worker who is assigned a map task reads the contents of the corresponding input split. It parses key/value pairs out of the input data and passes each pair to the user-defined Map function. Each Map function generates intermediate key/value pairs and stores them in memory.
4. Periodically the buffered pairs are written to local disk, partitioned into R regions by the partitioning function. The master keeps a record of the locations of R regions for forwarding locations to reduce workers.
5. When a reduce worker is notified by the master about these locations, it uses remote procedure calls to read the buffered data from the local disks of map workers. The reducer collects all intermediate data and sorts and groups them by intermediate keys, which is the sorting and shuffling stage on Hadoop.
6. Reduce worker call user defined Reduce function to process grouped intermediate data and write final results to the disk. The results are available in R output files.

In step 1, 4 and 6, overheads come from disk I/O. In step 5, overheads come from network communication and CPU resources usage of sorting and grouping. [2] shows that the time taken for data shuffling and sorting increases significantly as the number of data points increases from 50000 to 5 million, and the time taken to shuffle 50000 points was about 4 seconds which rose up to 30 seconds for 500000 points and 207 seconds for 5 million points. In those steps of one MapReduce job, time is consumed when CPUs and disks are not actually processing data, but storing, transferring or grouping data. In multiple sequential MapReduce jobs, much more time is consumed. So we definitely want to finish our task with as few MapReduce jobs as possible.

## (ii) K-means on MapReduce

K-means is one of the most popular and useful clustering method for large datasets. It aims to partition data into N dense groups. It starts with N initial centroids, groups data by their minimum distance to N centroids and averages data within each group to get new centroids and keeps doing until N centroids stop moving. Many methods are proposed to parallelize k-means [2, 3, 6, 7].

[3] proposes PKMeans. The pseudo code is as follows,

```
MapReduce_Job(centroids):

    Mapper_Func(key, instance):
        cluster_index = find_closest(instance, centroids)
        key' = cluster_index
        value' = instance
        output(key', value')
    End_Func

    Reducer_Func(key, array of instances):
        //where key is cluster_index from mapper
        new_centroid = average(array of instances)
        centroids[key] = new_centroid
        output(key, new_centroid)
    End_Func
End_Job

Main():
    centroids = initialcentroids
    previous_centroids = centroids
    while(distance(centroids, previous_centroids) > epsilon)
        // where epsilon is stop condition
        previous_centroids = centroids
        // keep creating and launching new MapReduce jobs
        JOB = MapReduce_Job(centroids)
        Launch(JOB)
    END_while
END_Main
```

Algorithm 1, PKMeans

Though PKMeans can have unlimited number of mappers, it can only have at most 1 reducer for each centroid, which makes it less efficient.

[2] proposes a similar algorithm. They notice that the overheads mainly come from the shuffling stage so they use a well designed combiner to reduce the amount of data that the mappers write and the reducers read.

[9] proposes a method using k-d tree to optimize the search for nearest centroids.

All of the above algorithms needs to launch M MapReduce jobs sequentially for M k-means iterations. How many iterations k-means needs? [4, 5] prove that the lower bound is $2^{\Omega(n)}$, where n is total number of data points. Thus using these algorithms for large dataset, we need exponentially many MapReduce jobs introducing huge overheads in total.

*(iii)    Our solution to reduce overheads*

We approach this problem differently, focusing on how to reduce the size of datasets. Our basic idea is to partition the whole dataset into several small representative subsets. Each reducer runs one complete k-means for each subset and generate 1 group of centroids, so k-means can be done in 1 single MapReduce job. M reducers generate M groups of intermediate centroids, which will be merged into 1 final group of centroids.

Two major advantages of this idea is that first, to do k-means, multiple cascading MapReduce jobs of PKMeans are converted into one parallel MapReduce job, and second, we can have much more parallel reducers than PKMeans. One problem we are faced with is how to get representative subsets. Random selection can't guarantee every sub region of the dataset has its representative points. To solve this problem, we come up with a method using k-d tree. In stead of using k-d tree to optimize the nearest computing [9, 10], we use it to partition datasets. K-d tree's building process and a proper selection strategy guarantee the representativeness of each subset. We also propose a parallel algorithm that builds k-d tree pretty fast using $O(\log(n))$ MapReduce jobs. Another problem is how to merge intermediate centroids. We propose two algorithms. One is hierarchical merging, the other is minimum average sum of square error (ASSE). Since we just do merging on M sets of centroids, which is a very small dataset, single machine is enough for running either of the 2 algorithms very fast. In general, our method includes 3 stages, dataset partitioning, parallel k-means clustering, and merging. The first 2 stages are parallel algorithms which can be done in $O(\log(n) + 1)$ MapReduce jobs. The last stage can be regarded as constant time. So with our method, the total number of MapReduce jobs needed is $O(\log(n) + 1)$ vs. PKMeans' $2^{\Omega(n)}$. So these are obvious advantages. Another indirect advantage is that every reducer runs a complete execution of a k-means independently, which means all optimization methods of single-machine k-means such as the method proposed by [9] can be used by each individual reducer to run faster. However, there are disadvantages as well. The major one is that each reducer only executes k-means on a small portion of the whole dataset, leading to loss of accuracy. Advantages and disadvantages will be fully discussed in Section 3.

The rest of paper is organized as follows. Section 2 describes our method in details. Section 3 presents our experiment results and discussion. Last section gives conclusion.

## 2. IPKMeans, the Improved Parallel k-means

Our method includes 3 stages, dataset partitioning, parallel k-means and single-machine merging. Let's assume we want K clusters. The dataset is first partitioned into M subsets and launch one MapReduce job with M reducers each of which is assigned a subset and runs 1 complete k-means on it resulting in K centroids. In total, M reducers generate K*M intermediate centroids, but we only want K centroids, so in the last stage, K*M centroids are merged into K final centroids.

*(i)    Dataset partitioning*

K-d tree is a space-partitioning data structure, for grouping nearest points. It uses a balanced binary tree to store points. Every leaf node stores equal number of nearest points in same sub region. Leaves are partitioned by hyper-planes. For 2-D datasets, building a k-d tree starts with splitting the dataset into 2 sub regions at median point along x axis, then for each sub region, independently do splitting at median point along y axis, then along x axis again, do splitting for all current sub regions, and keeps splitting along different directions until every sub region contains the number of points you want. For example, if you want 4 points in each final sub region, stop splitting when every sub region contains at most 4 points.

Now, K-d tree splits the dataset into R sub regions, each containing M points. To get representative subsets, 1 point is *picked* from M points of each sub region to form 1 subset, so each *subset* (not sub region!) contains R points, and the dataset is partitioned into M subsets.

How to pick 1 point from each sub region? Labelling is convenient for this purpose. Points in each sub region are labeled by numbers ranging from 1 to M. Points labeled 1 is picked by subset 1, points labeled 2 is picked by subset 2 and so on. There are 2 strategies to label points. One is to label points randomly, the other is to sort them along a certain axis and label them from leftmost to rightmost or topmost to bottommost. These variations will be discussed later.

A parallel MapReduce algorithm is proposed to build k-d tree (Algorithm 2).

```
MapReduce_Job(subregion_points_count)
      /* Before job launching, add suffix "0" to each instance as sub region ID
to indicate that they are initially in the same region. When launching the job,
mapper will receive both suffix and data instance as value. subre-
gion_points_count is used to count how many points are in one leaf node.*/
      Mapper_Func(key, value)
            [instance, suffix] = parse(value)
            subregion_id = suffix
            key' = subregion_id
            value' = instance
            output(key', value')
      END_Func
      Reducer_Func(key, array of values)
            subregion_points_count = length(subregion_points_count)
            subregion_id = key
            instance_array = array of values
            [subarray1, subarray2] = split_by_median(instance_array)
            for every instance in subarray1
                  key' = id of instance
                  new_suffix = suffix.append('0')
                  value' = instance and new_suffix
                  output(key', value')
            END_for
            For every instance in subarray2
                  key' = id of instance
                  new_suffix = suffix.append('1')
                  value' = instance and new_suffix
                  output(key', value')
            END_for
      END_Func
END_Job

Main()
      subregion_points_count = INFINITY
      leaf_node_capacity = 6 // number chosen by user
      while(subregion_points_count > subregion_points_count)
            JOB = MapReduce_Job(subregion_points_count)
            Launch(JOB)
      END_while
END_Main
```

Algorithm 2, building k-d tree

After building the k-d tree, using its outputs, another MapReduce job is launched to generate representative subsets (Algorithm 3).

```
MapReduce_Job()
      /* send outputs of last step to mapper*/
      Mapper_Func(key, value)
            [instance, suffix] = parse(value)
            subregion_id = suffix
            key' = subregion_id
            value' = instance
            output(key', value')
      END_Func
      Reducer_Func(key, array of values)
            len = length(array of values)
            instance_array = array of values
            sort instance_array along a certain axis
            random_labels = generate random permutation for 1, 2, 3, … len
            for i = 1 to len
                  // below are 2 different options for setting key
                  key' = random_labels[i]   // random labeling
                  key' = i             // labeling along a certain direction
                  value' = instance_array[i]
                  output(key', value')
            END_for
      END_Func
END_Job
```

Algorithm 3, k-d tree sub region labeling for generating subsets

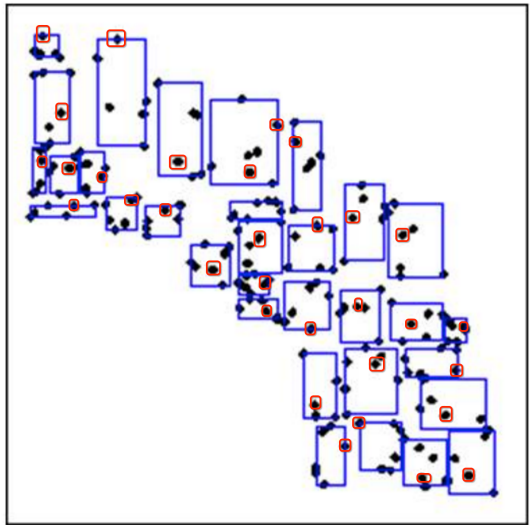

Figure 1, labeling data points. Points in same subset is showed by small red rectangles.

Assuming each sub region contains at most M instances, after running Algorithm 3, each data instance gets a label ranging from 1 to M randomly or along a certain direction. Instances with same label are considered as one subset, so M subsets are generated. The more subsets are generated; the more parallel reducers the following k-means can use.

### (ii) k-means done in one MapReduce job

Instead of launching multiple MapReduce jobs sequentially in a while loop, only one MapReduce job is started to do k-means as in Algorithm 4. One subset is mapped to one reducer. Each reducer starts a k-means on its assigned subset and keeps running all the way until k-means converges. One problem is that each k-means only knows a portion of the whole dataset, so how accurate of this method can be? In the next section, several experiments are conducted to look into this problem.

```
MapReduce_Job()
      /* send ouputs of last step to mapper*/
      Mapper_Func(key, value)
            [instance, label] = parse(value)
            subset_id = label
            key' = subset_id
            value' = instance
            output(key', value')
      END_Func
      Reducer_Func(key, array of values)
            instance_array = array of values
            [centroids, average SSE]
                  = run k-means for instance_array until kmeans converges
            key' = key
            value' = centroids with average SSE (sum of square error)
            ouput(key', value')
      END_Func
END_Job
```
Algorithm 4, k-means on subsets done in 1 MapReduce job

### (iii) Merging

The input of merging is the outputs of last step, i.e. K*M intermediate centroids, where M is the number of reducers, and K is the number of centroids generated by each reducer.

We propose 2 methods to do merging,

(a) Hierarchical merging (Algorithm 5)

We keep replacing the closest 2 points in the array of centroids by their midpoint until the number of centroids equal the number of clusters we want. The time complexity is $O(N^3)$. In usual cases, N is small, so we can treat this as constant time. If necessary, this algorithm can be optimized using GPU or MapReduce.

```
Merging(centroids, number_of_cluster)
      while(centroids.size() > number_of_cluster)
            [point1, point2] = find closest 2 points in centroids
            centroids.remove(point1)
            centroids.remove(point2)
            new_point = compute midpoint for point1 and point2
            centroids.add(new_point)
      END_while
END_Merging
```
Algorithm 5, hierarchical Merging

(b) Minimum ASSE (average SSE)

(ii) generates intermediate centroids with the corresponding ASSE of their subsets. ASSE is the average distance from data points to its nearest centroid.

In k-d tree preprocessing stage, we choose labeling along a certain axis to generate subset (Algorithm 3). It's reasonable to assume that the point near the center of its sub region is more representative. Therefore, hopefully the more representative a subset is, the lower ASSE it has.

The subset with minimum ASSE can be found by doing k-means for every subset and computing the ASSE. This algorithm is very straightforward. First, all intermediate centroids are traversed and the ones with minimum ASSE are picked as final centroids. The time complexity is O(N), where N is the number of intermediate centroids. Though this algorithm is very simple, experiments show that it's more robust and reliable than hierarchical merging.

(iv) Method overview

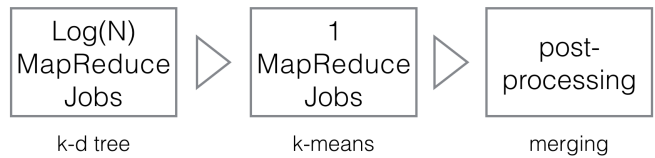

Figure 2, IPKMeans

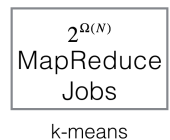

Figure 3, PKMeans

Figure 2, 3 show that IPKMeans needs log(N) + 1 MapReduce jobs and PKMeans needs $2^{\Omega(N)}$ MapReduce jobs [4, 5], where N is the total number of points. The number of MapReduce jobs is reduced from $2^{\Omega(N)}$ to log(N) + 1. The following experiments show that overheads are greatly reduced.

## 3. Results and Discussion

(i) Environment

2.9 GHz Intel Core i5, Mac OS X, Hadoop 1.2.1 in multithreaded debug mode.

(ii) Variations of IPKMeans

As we mentioned before, there are 3 options to do dataset partitioning,
1) k-d tree splitting + random labeling within sub regions.
2) k-d tree splitting + labeling along a certain axis within sub regions.
3) no k-d tree + random partition.
There are 2 options to do merging,
a) hierarchical merging.
b) Minimum ASSE.
Our experiments show that the combination of 2) and b) outperforms other combinations.
(iii), (iv) and (vi) are all conducted with 2) and b). In (v), other combinations are discussed.

### (iii) Performance comparison of IPKMeans and PKMeans

The dataset used includes 3000 of Gaussian distributed 2-D points and 5 groups of centroids (Figure 4). (iii), (iv) and (v) use this dataset. (vi) uses a dataset with more points.

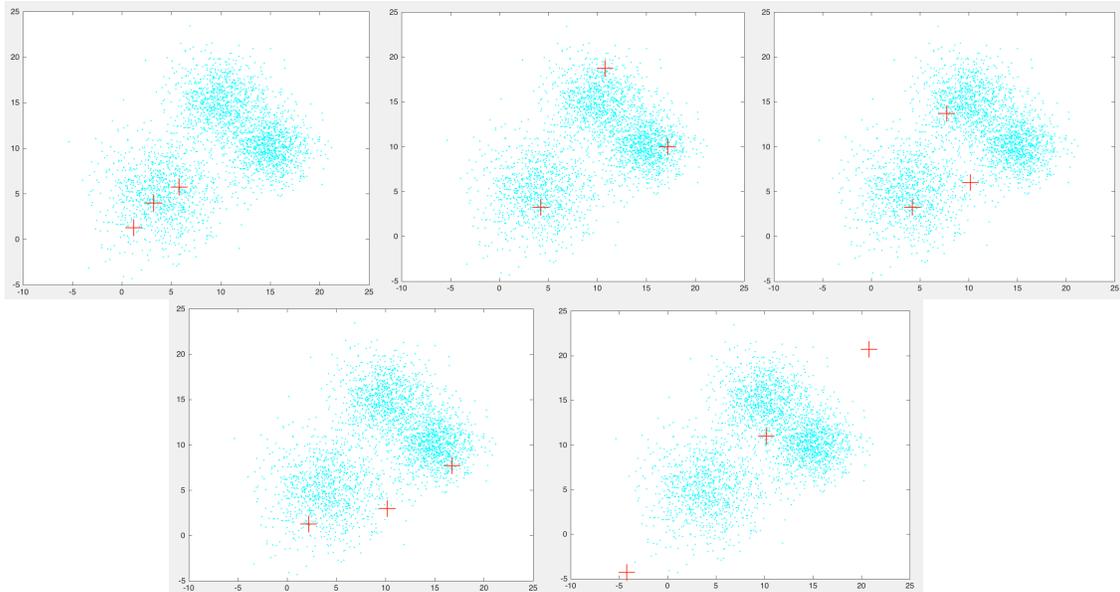

Figure 4, the dataset with 3000 points and 5 groups of centroids marked by '+'

In this bunch of experiments, both algorithms use same default settings of Hadoop debug mode and 5 experiments are conducted on same dataset and 5 different groups of initial centroids with each experiment using same initial centroids. IPKMeans splits dataset into 6 subsets using parallel k-d tree with labeling along a certain direction, i.e. 2) + b) in (ii).

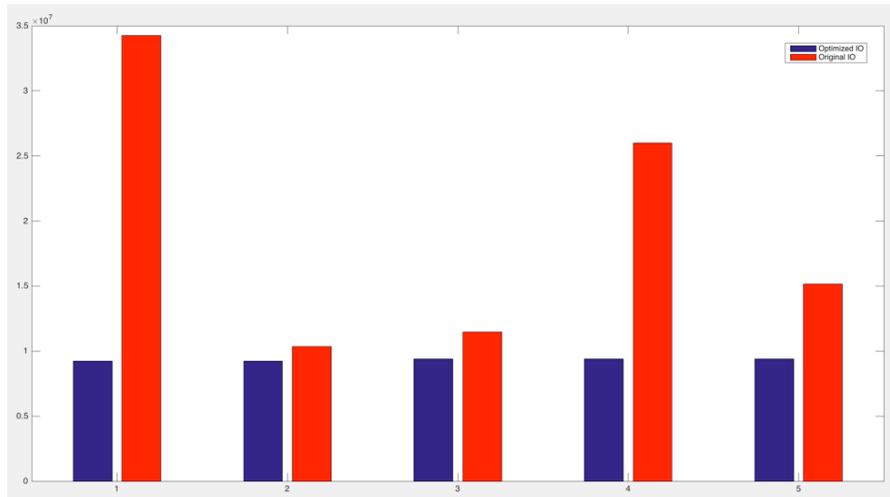

Figure 5, disk reading and writing of IPKMeans (blue/left bars) vs. PKMeans (red/right bars) in bytes

As Figure 5 shows, IPKMeans has up to 2/3 lower I/O overheads than PKMeans.

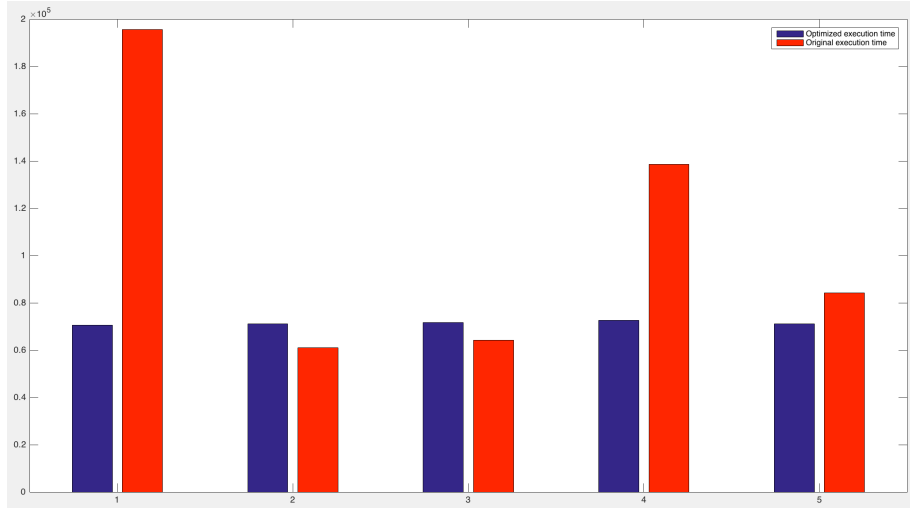

Figure 6, execution time of IPKMeans (blue/left bars) vs. PKMeans (red/right bars) in ms.

As Figure 6 shows, IPKMeans saves up to 2/3 execution time than PKMeans. In experiment 2 and 3, IPKMeans is slightly slower than PKMeans. That was because centroids were so well chosen in these 2 experiments that PKMeans converges after only 5-8 iterations and meanwhile k-d tree could not make the following k-means faster enough to outperform PKMeans. Another reason is that IPKMeans only used 6 subsets, and it could use more to be faster.

In terms of the comparison of clustering accuracy, SSE is a good metric.

| Experiments | 1 | 2 | 3 | 4 | 5 |
|---|---|---|---|---|---|
| SSE of PKMeans | 3.4817e+04 | 3.4817e+04 | 3.4817e+04 | 3.4817e+04 | 3.4817e+04 |
| SSE of IPKMeans | 3.4843e+04 | 3.4843e+04 | 3.4841e+04 | 3.4843e+04 | 3.4841e+04 |

Table 1, SSE of IPKMeans vs. PKMeans

Table 1 shows that the SSEs of PKMeans of IPKMeans are very close indicating close clustering results.

*(iv)    IPKMeans with more parallel reducers*

The number of parallel reducers equals the number of subsets. So more subsets indicate more reducers and higher parallel efficiency, but with more subsets, each subset contains fewer data points, making each subset less representative leading to worse clustering accuracy.

5 experiments are conducted on dataset 1 with 6, 11, 23, 46, 93 reducers.

| Experiments | 1 | 2 | 3 | 4 | 5 |
|---|---|---|---|---|---|
| Reducers | 6 | 11 | 23 | 46 | 93 |
| Time of Exe.(ms) | 71241 | 65434 | 61486 | 50297 | 45318 |
| SSE | 34842.58 | 34873.53 | 34947.37 | 35524.56 | 37107.97 |

Table 2, experiments of different numbers of reducers.

As more parallel reducers are used, IPKMeans runs faster. The reasons are twofold. First, k-d tree needs fewer iterations and second. Second, more reducers are working in parallel. With 93 reducers, IPKMeans uses nearly 40% less execution time than that with 6 reducers. As Table 2 shows, SSE slightly increases 6.5% from 34842.58 to 37107.97, indicating that we gain 40% speed-up losing only 6.5% accuracy.

Additionally, the multi-threaded debug mode of Hadoop can't run such many reducers in real parallel, so theoretically, a cluster of 100 nodes should give much better performance.

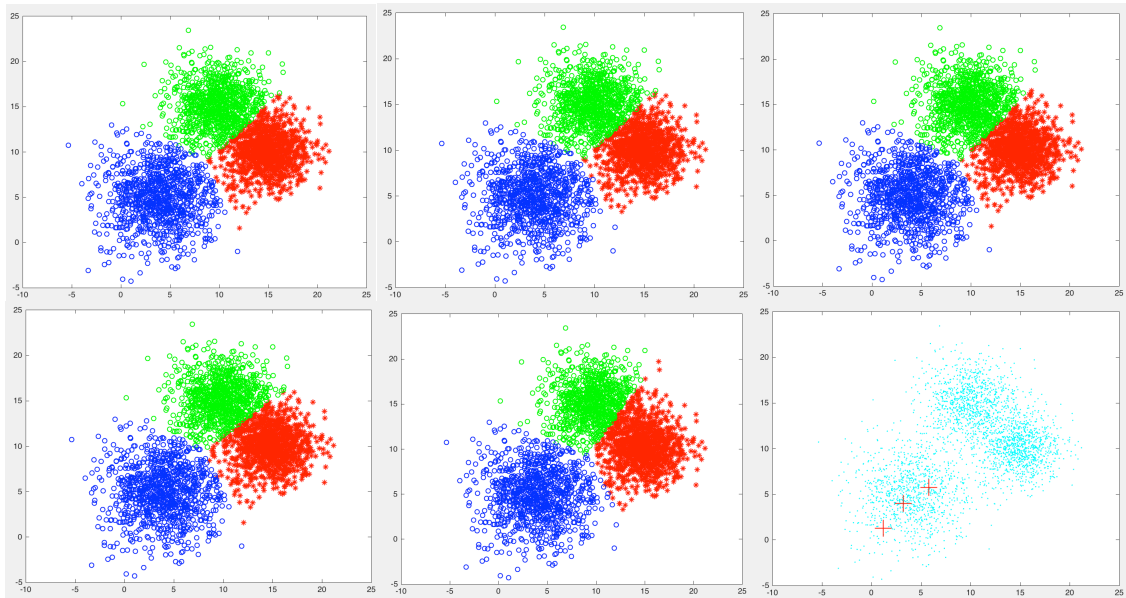

Figure 7, clustering results of experiments 1-5 with 6, 11, 23, 46, 93 reducers. The last image shows initial centroids marked by '+'.

The current dataset of 3000 points is too small for more reducers. 3000 points are split into 93 groups assigned to 93 reducers. Each reducer only processes about 30 points. 30 points may not be representative enough to generate reliable clustering. With a larger dataset, each reducer receives more data points so we can have more parallel reducers to achieve higher accuracy and efficiency.

*(v)    Discussion of different variations of IPKMeans*

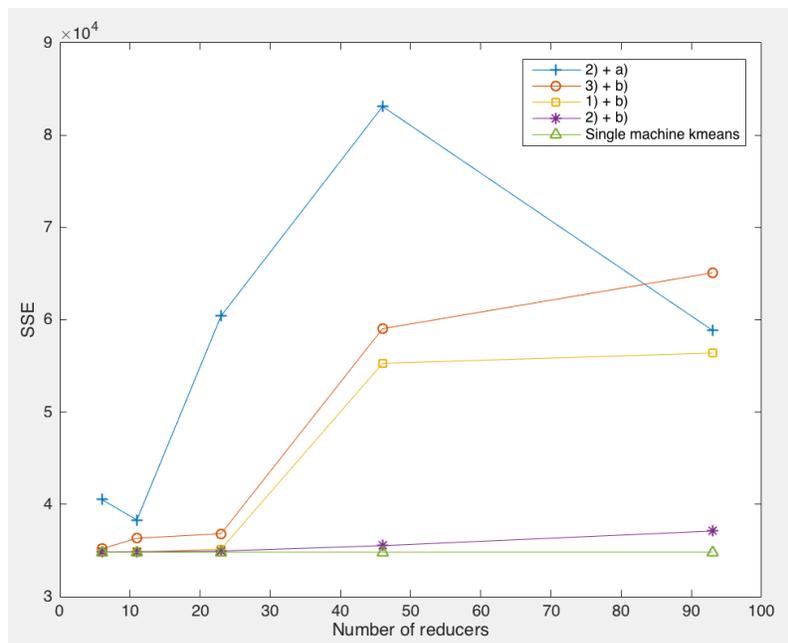

Figure 8, SSE of different combinations in (ii). Y axis is SSE. X axis is number of reducers.

Experiments are conducted for many combinations in (ii) using 6, 11, 23, 46, and 93 reducers.

According to Figure 8, the performances of different combinations are arranged from worst to best, 2) + a) < 3) + b) < 1) + b) < 2) + b). The single machine k-means clustering serves as benchmark. With hierarchical merging, good centroids may be merged by bad centroids, so the result is not stable. 1) + b) is slightly better than 3) + b). 3) uses global random partitioning without k-d tree. When the dataset is partitioned into more subsets and each subset contains fewer points, random partitioning can't guarantee any representativeness. Chances are that no subset is representative. 1) uses random labeling in sub regions of k-d tree, which is slightly better than 3), because it guarantees every sub region has at least 1 representative point. 2) + b) is the best. Intuitively, the reason is that points in same region are labeled along a certain axis, so at least one subset fitting with initial centroids well can be found to be representative enough to make good clustering.

(vi)    *IPKMeans on larger datasets*

A few more experiments are conducted on a larger dataset with 15000 points, 4 clusters and fixed initial centroids. With 58, 117, 234, 468, 937 parallel reducers, IPKMeans gives good results (Table 3 and Figure 9). There is no need to do experiments with less than 58 reducers, because accuracy always increases with fewer reducers (one-reducer IPKMeans equals to single-machine k-means).

The SSE of single machine k-means is 1.3178e+05. The time taken by PKMeans is 83551 ms.

| Experiments | 1 | 2 | 3 | 4 | 5 |
|---|---|---|---|---|---|
| Reducers | 58 | 117 | 234 | 468 | 937 |
| SSE | 1.3208e+05 | 1.3245e+05 | 1.3275e+05 | 1.3791e+05 | 1.6950e+05 |
| Time of Exe.(ms) | 65510 | 60183 | 52052 | 45413 | 39294 |

Table 3, experiments on 15000 points and 4 clusters

With 937 reducers, each reducer is assigned a subset with only 15 points. Amazingly, in such a difficult case intended for pushing IPKMeans to the limit, it still gives a not so bad result. The results of first 4 experiments are almost perfect.

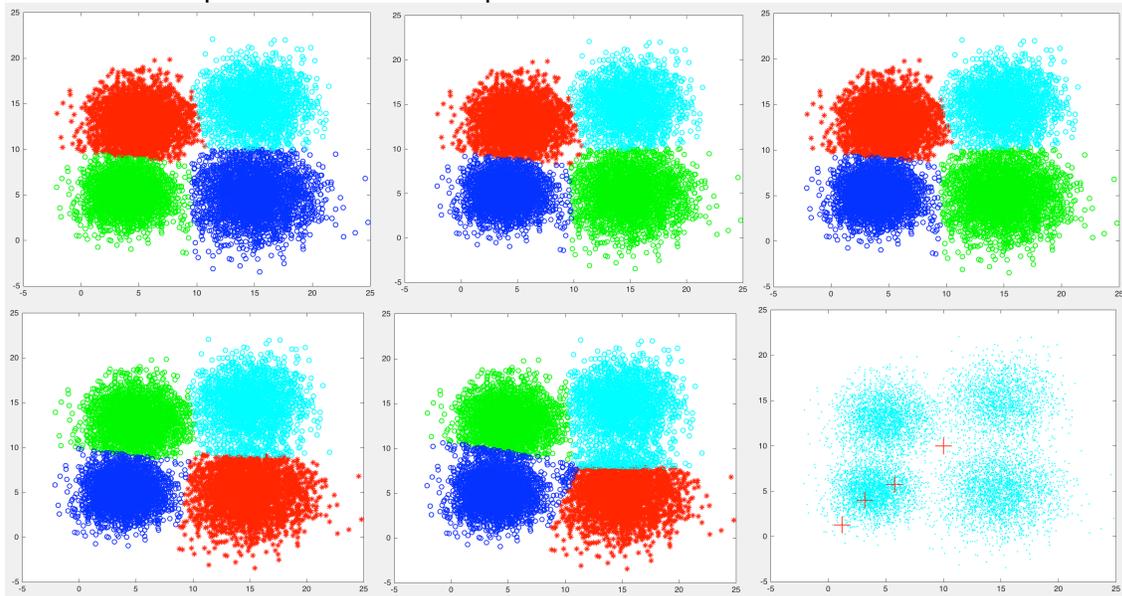

Figure 9, results of experiments 1-5. The last image shows initial centroids marked by '+'.

The time of execution in Table 3 can't represent the efficiency of IPKMeans running in real world. In real parallel environment, IPKMeans can be much faster.

## 4. Conclusion

We propose a new parallelization method for k-means, IPKMeans. In terms of efficiency, at least on our datasets, all experiments with different settings show that IPKMeans outperforms all other parallelization methods. With larger datasets, more reducers can be used to gain more speed-up while still keeping almost the same accuracy.